# Ultra-high bandwidth quantum secured data transmission

James F. Dynes[1,*], Winci W-S. Tam[1], Alan Plews[1], Bernd Fröhlich[1], Andrew W. Sharpe[1], Marco Lucamarini[1], Zhiliang Yuan[1], Christian Radig[2], Andrew Straw[3], Tim Edwards[3] and Andrew J. Shields[1]

[1]Toshiba Research Europe Ltd, 208 Cambridge Science Park, Cambridge CB4 0GZ, UK. [2]ADVA Optical Networking SE, Maerzenquelle 1-3, 98617 Meiningen, Germany. [3]ADVA Optical Networking, York, Advantage House, Tribune Way, Clifton Moore, Tribune Way, York, YO30 4RY, UK.

*james.dynes@crl.toshiba.co.uk

**Quantum key distribution (QKD) provides an attractive means for securing communications in optical fibre networks. However, deployment of the technology has been hampered by the frequent need for dedicated dark fibres to segregate the very weak quantum signals from conventional traffic. Up until now the coexistence of QKD with data has been limited to bandwidths that are orders of magnitude below those commonly employed in fibre optic communication networks. Using an optimised wavelength divisional multiplexing scheme, we transport QKD and the prevalent 100 Gb/s data format in the forward direction over the same fibre for the first time. We show a full quantum encryption system operating with a bandwidth of 200 Gb/s over a 100 km fibre. Exploring the ultimate limits of the technology by experimental measurements of the Raman noise, we demonstrate it is feasible to combine QKD with 10 Tb/s of data over a 50 km link. These results suggest it will be possible to integrate QKD and other quantum photonic technologies into high

**bandwidth data communication infrastructures, thereby allowing their widespread deployment.** Most demonstrations of QKD[1-2] to date have used dark fibre, in which the quantum signals are transmitted separately from the conventional data[3-5]. However, these additional dedicated links are not always available, and even when they are, can be prohibitively expensive for the vast majority of applications. Thus it is imperative for cost effective deployment of the technology, that QKD signals share the same fibre as conventional data[6]. This is very challenging as the data lasers are typically 8 orders of magnitude stronger and produce a broad background spectrum of scattered light that can overwhelm the much weaker quantum signals[7] and experimental demonstrations to date have been limited to data bandwidths of a few tens of Gb/s, orders of magnitude less than the bandwidths commonly employed on links in fibre optic communications[7-12].

In this report we establish that quantum communication is compatible with high bandwidth optical networking infrastructure. We demonstrate a quantum encryption system, combining QKD with Mb/s key rates and encrypted data transport with a bandwidth of 200 Gb/s on the same fibre. We combine QKD with dual polarisation quadrature phase shift keying (DP-QPSK) for the data channels for the first time. The system operates for fibres up to 101 km in length, sufficient for nearly all links found in metropolitan area networks. Furthermore, we explore the effect of even higher data bandwidths, demonstrating that multiplexing QKD and up to 10 Tb/s of data is feasible for fibre lengths up to 50 km.

Figure 1 illustrates the combined quantum and classical network layout with a single optical fibre link used to transmit traffic from both the QKD system (see Methods for details on the QKD protocol) and high speed encryptors (HSE). The quantum signals are assigned to a single wavelength (1547.72 nm) on the International Telecommunication Union (ITU) Dense Wavelength Division Multiplexing (DWDM)[13] grid corresponding to channel number 37. We use fibre optic wavelength multiplexers to combine the QKD signal with multiple data channels on the DWDM grid around 1530 nm.

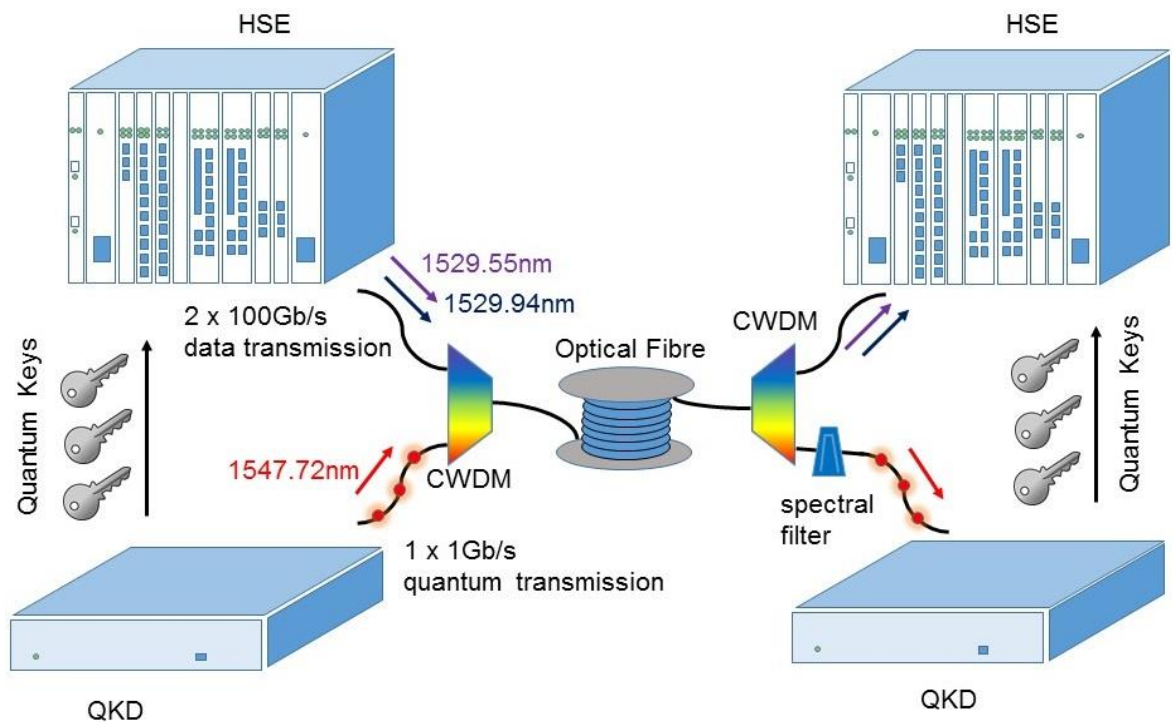

**Figure 1 Quantum secured system for ultra-high data bandwidth encryption.** A high speed quantum key distribution (QKD) system is combined with a high speed classical data encryptor (HSE). Details of the QKD system can be found in Ref.[14] and the working principle of the HSE can be found in Ref.[15]. The multiple wavelengths from the HSE are wavelength multiplexed using a dense wavelength division multiplexer (DWDM) before multiplexing with the quantum signals using a coarse wavelength division multiplexer (CWDM). Quantum signals on the receiver side are spectrally filtered using either a 100GHz or 25GHz spectral filter. Quantum keys generated by the QKD system are pushed symmetrically over Ethernet to the respective HSE's. The quantum keys are used to encrypt the 2 × 100 Gb/s data traffic using the AES algorithm in real time.

The data transport layer features HSE units with two 100G Advanced Encryption Standard (AES) encryptor line cards operating DP-QPSK with wavelengths (ITU grid channel numbers) 1529.55 nm (60) and 1529.94 nm (60.5) respectively. We chose to adopt a DP-QPSK scheme[15] as phase shift keying (PSK) permits a better optical signal to noise ratio of approximately 3dB than simple on-off keying (OOK) schemes for the same optical receiving power.[15-16] We also remark the data transmission capacity of DP-QPSK is four times higher than that of OOK. We use wavelength divisional multiplexing to combine the data and quantum signals on a single fibre pair, with all forward directed traffic in the first fibre and the backward directed traffic in the second. This follows the usual practice of using a pair of fibres for bi-directional communication[17]. As well as conforming to the standard approach, this helps reduce noise on the quantum channel for fibre distances greater than about 25 km as Raman scattering due to strong data signals in the forward direction weakens due to fibre attenuation[8].

We have adapted the built-in Advanced Encryption Standard (AES) encryption on the 100G line cards to accept symmetrically pushed quantum keys. This replaces the conventional public key exchange usually required to perform AES encryption. 512 bits are extracted from the pushed key and half of these (i.e. 256 bits) are used by each 100G line card for AES encryption in counter mode[18].

To minimise the deleterious effect of Raman scattering into the quantum channel[7-12], we employ a combination of spectral, temporal and optical power control. We reduce the launch power of the transmitters into the fibre by amplifying the data signals at the receiver side using an Erbium doped fibre amplifier. The launch power was chosen to maintain error free data transport after forward error correction. The value depends on the fibre link loss; for example at a fibre distance of 50 km we set the launch power to -25.5 dBm (2.8 µW).

Raman scattering into the quantum channel is minimised by an optimised combination of spectral and temporal filtering. Temporal filtering is accomplished using self-differencing[19] operation of the avalanche photo-diodes (APDs) in the quantum receiver. When operated in self-differencing mode, the APDs can both detect very weak avalanches[19] as well exhibit low avalanche signal evolution[20]. This means only avalanches occurring at the start of the clock edge contribute to photon counting which results in a short active on-time. Therefore single photon detectors based on self-differencing APDs exhibit inherent temporal filtering. Temporal filtering results in an effective active on-time of around 125 ps, when the single photon detectors are gated at a repetition rate of 1 GHz. This means ~ 88% of Raman scattered photons are automatically rejected by temporal filtering through the self-differencing detection technique[21]. The quantum channel is also filtered spectrally using thin film filters with a nominal spectral width of either 100 GHz or 25 GHz (in the latter case, the measured full width at half maximum was 15 GHz).

Figure 2 shows the dependence of the secure bit rate as a function of optical fibre distance up to 101 km in the presence of 2 × 100 Gbps forward directed encrypted data for the two types of filters. For the 100 GHz filter, a secure bit rate of 1.9 Mbps (1.2 Mbps) is observed for fibre distances of 35.5 km (50.5 km), sufficient to generate an AES-256 encryption key every 200 $\mu$s. The corresponding fibre losses in these cases were 6.8 dB and 9.6dB respectively.

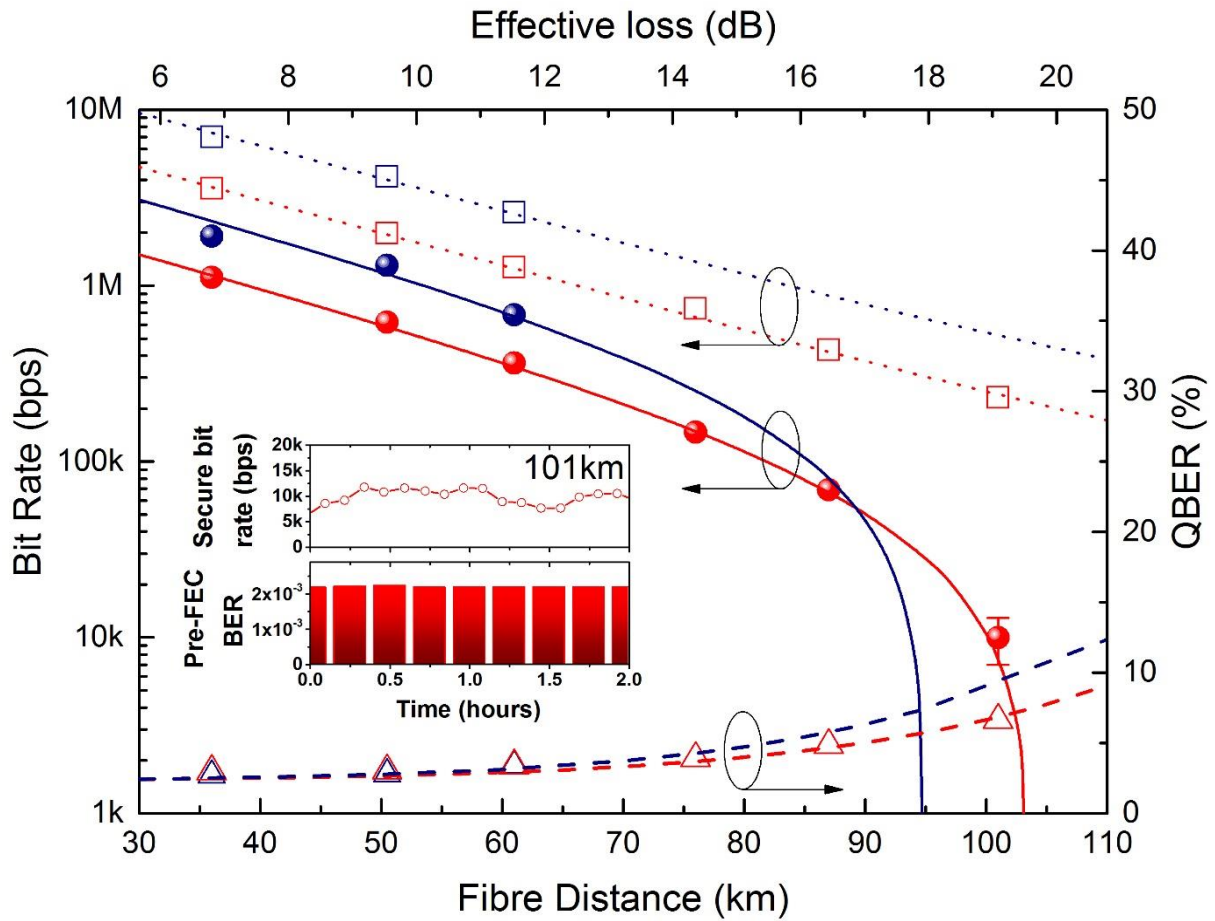

**Figure 2 Fibre distance dependence.** Experimental secure bit rate as a function of fibre distance in the presence of 2 × 100G forward directed classical data traffic over the same fibre (see Methods for details on the QKD protocol). Data is reported for two different filter widths, 100 GHz (blue circles) and 25 GHz (red circles). Corresponding sifted bit rates and quantum bit error rates are displayed as squares and triangles respectively. The solid and dashed lines are calculated using numerical simulation (see refs 8, 11 & 15 for more details). The simulation fully accounts for Raman forward scattering by the classical data traffic into the quantum channel. Error bar is two standard deviations. Effective fibre losses are shown on the top axis. Inset: Secure bit rate and data bit error rate (before forward error correction) for a duration of 2 hours at a fibre length of 101 km.

Simulation of the secure bit rate with the 100 GHz filter shows that positive secure bit rates are not possible at 100 km (blue line, Figure 2). To elongate the range of the quantum encryption system, we exchange the 100 GHz filter in the quantum receiver for a 25 GHz filter (red circles). The effect of the narrower filter is better Raman noise rejection, as evidenced by the lower QBER at longer distances (red triangles).

Although the secure bit rates at shorter distances are now lower due to the higher loss of the 25 GHz filter, the reach is extended to 101 km (equivalent loss of 19.1dB) with a corresponding secure bit rate of 10 kbps. Continuous operation over 2 hours is also shown in the top panel of the inset, Figure 2. This secure bit rate is still sufficient to refresh the AES encryption key every 25 ms.

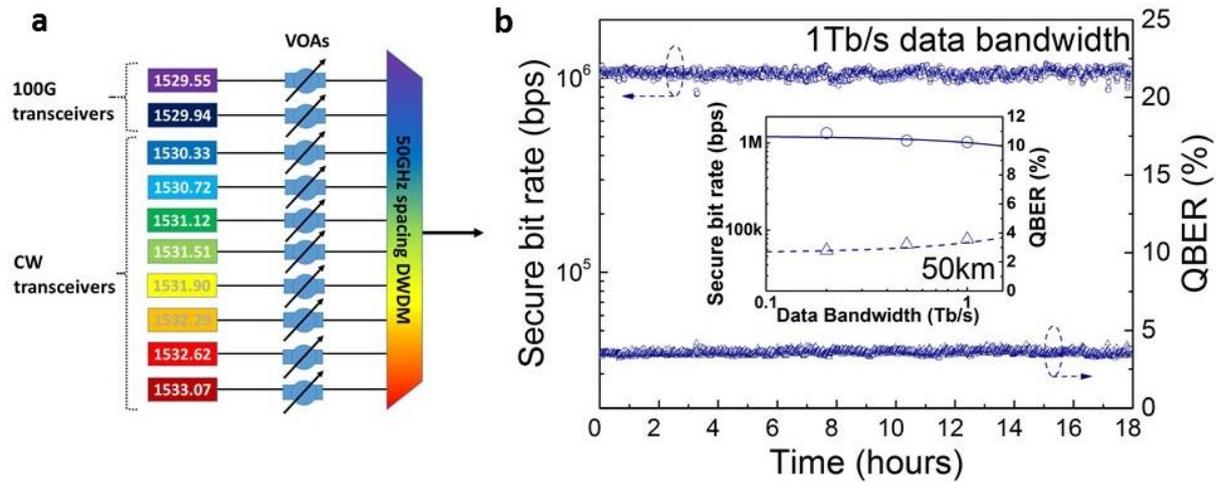

**Figure 3 High bandwidth experimental simulation. a**, Eight continuous wave (CW) optical transceivers are wavelength multiplexed with the original two 100G transceivers. Each transceiver operates on a separate wavelength of the 50GHz DWDM grid spanning 1529.55 nm → 1533.07 nm. The optical launch power of each transceiver is set using a variable optical attenuator (VOA) to -25.5 dBm. **b**, Secure bit rate as a function of time for 18 hours of continuous operation in the presence of 10 classical data channels representing 1 Tbps of data traffic over the same fibre (blue circles). The corresponding quantum bit error rate (blue triangles) is also shown. Inset: Data bandwidth dependence of the secure bit rate and QBER from 0.1 → 1 Tbps. Solid and dashed lines are results from numerical simulation.

We now explore the maximum data bandwidth that can co-exist with QKD signals. For this study, we populate the optical link with additional CW lasers operating at separate wavelengths to simulate Raman noise created by the additional data traffic. The arrangement is shown in Fig. 3a. Eight supplementary data lasers on a 50 GHz grid are multiplexed together with the original 100G transceivers using the DWDM combiner over the wavelength range 1530.33nm – 1533.07nm. A fibre with a length of 50km and an average loss of 0.19dB/km is used. This distance was chosen since it is greater than the reach for typical 100G signal deployments, a reach specified to be transported across a link of up to 40km as per the IEEE standard[22] (currently no standard exists for 1 Tbps or 10 Tbps data). We set the launch powers of the additional data lasers to the same level as the 100G transceivers; in this case -25.5 dBm (2.8 µW). As before a 100 GHz optical filter is used to spectrally isolate the quantum signals at the quantum receiver side.

Raising the power of the data lasers in the optical fibre link is expected to increase the Raman scatter and thereby deteriorate the QKD performance. However, the observed secure bit rate with all 10 transceivers switched on shows very little reduction. A secure bit rate of > 1 Mbps is achieved over a duration of 18 hours, Figure 3b, blue circles. This demonstrates that QKD can co-exist with 10 × 100G data channels, corresponding to an aggregate bandwidth of 1 Tbps.

To investigate the ultimate bandwidth that can be combined with QKD, we increased the launch power of the 10 data lasers. This simulates the effect of adding additional data channels, provided that the noise in the quantum channel varies linearly with the data laser power. Furthermore, since the 100G channels are launched at a wavelength around 1530nm, this reflects the worst case of Raman scattering into a quantum channel wavelength of ~ 1550nm for the C-band wavelength range between 1530 and 1565nm, since the Raman scattering coefficient for 1530nm is the highest. We also note the 25GHz filter can in principle have very high isolation – thus the Raman scattering effect into the quantum channel of adjacent classical channels would be the same as for non-adjacent classical channels. In the following experiment we ensure that the total optical power is < 0 dBm so that any non-linear effects in the fibre are negligible[12].

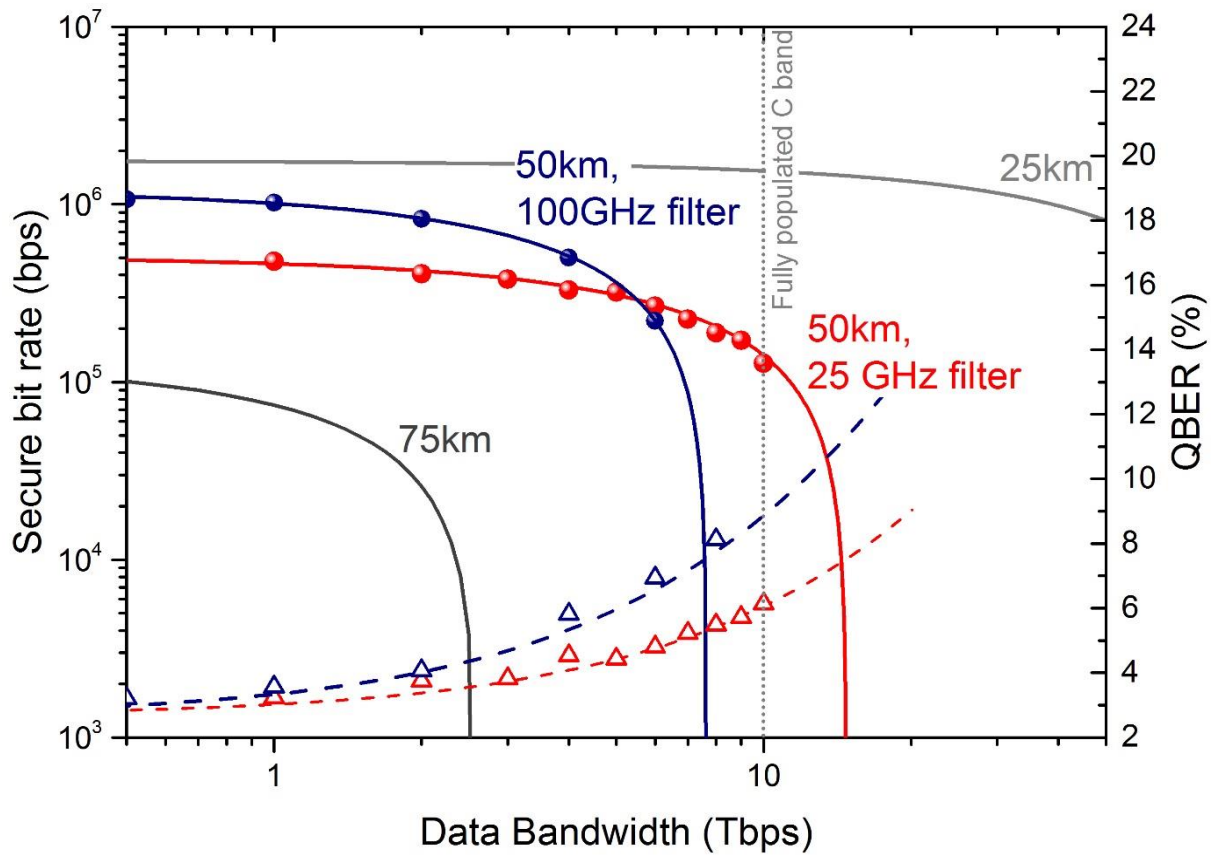

**Figure 4 Secure key bit rate data bandwidth dependence.** Experimental secure bit rate as a function of data bandwidth simulated by increasing the launch power of 10 data lasers. Experimental secure bit rate using a 100 GHz filter in the quantum receiver (blue circles), extending the results reported in the inset of Figure 3. Experimental secure bit rate data using a 25 GHz filter in the quantum receiver (red circles). Blue and red triangles are the corresponding QBERs for both experiments. Solid and dashed lines are results from numerical simulations. Also shown are secure bit rate numerical simulations using a 25 GHz filter in the quantum receiver for fibre distances of 25 km and 75 km (grey lines).

We first look at extending the data bandwidth when using a 100GHz filter in the quantum receiver. Figure 4 shows the secure bit rate dependence on the experimentally simulated data bandwidth for this case (blue circles). Notice that QKD with a finite bit rate is possible for bandwidths up to 6 Tb/s. At these high simulated data bandwidths almost all the contribution to the QBER (blue triangles) is due to Raman scattering from the data lasers. These results are improved further by use of a 25 GHz filter in the quantum receiver (red circles). In this case the data bandwidth can reach 10 Tb/s, along with a QKD key rate of 139 kb/s.

Theoretical simulations, shown as solid lines in Figure 4, are plotted for three separate distances, 25 km, 50 km and 75 km. The longer distance of 75 km illustrates positive secure bit rates even beyond a simulated bandwidth of 1 Tb/s. The shorter distance of 25 km indicates the secure bit rate remains above 1 Mbps up to and beyond a simulated data bandwidth of 10 Tb/s. The experimental and theoretical simulated results far exceed those reported previously in the literature[7-12] both in terms of data bandwidth and secure bit rate supported.

Our results suggest that QKD can co-exist with very high volumes of data, in excess of 10 Tb/s for fibres up to 50 km, transmitted simultaneously on the same fibre. This bandwidth is equivalent to one hundred 100 Gb/s data channels, sufficient to fill most of the C-band. The secure key rate at 10 Tb/s is ~139 kb/s, which is adequate for sub-second (~0.2 s) refresh of the AES encryption key on each of the one hundred 100G data channels. Our approach is particularly appropriate to applications such as off-site backup or data center interconnections, where many channels of homogenous traffic might naturally share the same transmission hardware over a point-to-point link and the optical launch power can be easily controlled. In the future the bandwidth of data multiplexed with QKD may be further increased by using multi-mode or multi-core optical fibre[23-25].

## Methods

**Quantum key distribution system.** The QKD system is composed of two 19 inch rack units running a phase encoded BB84 protocol with efficient basis selection and decoy states in the finite key size regime[14]. The QKD transmitter uses pulses at a wavelength of 1547.72 nm and an average intensity of 0.4 photons/pulse for the signal states. Two decoy states are used with intensities of 0.1 and $7\times10^{-3}$ photons/pulse. Phase encoding and decoding is achieved by the use of asymmetric Mach-Zehnder interferometers and the basis probabilities are set to the biased case of 31/32 and 1/32 for the majority and minority bases respectively. An optical MHz clock signal is transmitted over the same fibre for synchronisation of the QKD transmitter and receiver. The QKD receiver contains two InGaAs/InP avalanche photodiodes operating in self-differencing mode[3] clocked at 1 GHz with single photon efficiencies of 22.5% and dark count probabilities of $4.5\times10^{-6}$. Simulations of the secure bit rates and QBERs in the presence of classical channels used experimentally determined scattering coefficients for each of the classical launched wavelengths.

**High Speed Encryptors.** The HSE classical transport layer are two ADVA FSP3000 19 inch rack units which house two 100G AES encryptor line cards each containing a C form-factor pluggable (CFP) transceiver operating quadrature phase shift keying and dual polarization (DP-QPSK) with wavelengths (ITU grid channel numbers) 1529.55 nm (60) and 1529.94 nm (60.5) respectively. A 96 channel DWDM multiplexer/demultiplexer filter featuring an insertion loss of approximately 5 dB was used to combine the two HSE wavelengths, before mixing them with the quantum traffic by virtue of the CWDM multiplexing scheme. The HSE 100G AES line cards have built in forward error correction (FEC). To maintain post-FEC error free operation (BER ≤ $10^{-15}$), it is important that the pre-FEC bit error rate (BER) does not exceed the FEC threshold of $1.9 \times 10^{-2}$. We studied the pre-FEC BER at a fibre distance of 101 km since this had the highest optical loss of any distance and any aggravating effects, such as degradation in the optical signal to noise ratio, would be strongest. As can be seen from the inset in Figure 2, the pre-FEC BER over a continuous duration of 2 hours for a fibre distance of 101 km was very stable, at an average value of $2.2 \times 10^{-3}$ which is almost a factor of ten less than the pre-FEC threshold. The QKD generated keys at both the QKD transmitter and QKD receiver sides are pushed over their local communication interfaces to their respective high speed encryptors. These keys are then used by replacing the usual publically exchanged keys for the AES encryption algorithm to create a symmetrically enabled AES encryption link.

**Multiplexing scheme.** We choose to take advantage of the naturally low insertion loss of around 1 dB[7] for Coarse Wavelength Division Multiplexing (CWDM) to combine the quantum communication with the classical communication. Although CWDM technology is designed to support up to only 8 usable wavelength channels (around the low loss fibre wavelength of 1550nm), it can be adapted to work in conjunction with DWDM architecture. The CWDM channel bands are around 18 nm wide indicating they can support up to 22 DWDM 100 GHz spaced channels or 44 DWDM 50 GHz spaced channels in each band[26]. In fact, using just two CWDM wavelength bands (1530 nm and 1550 nm), almost the entire 50 GHz DWDM spaced ITU C-band can be populated. All forward directed signals: quantum signal, the QKD synchronization signal, QKD reconciliation data signal and classical channels for ultra-high bandwidth secured data transmission are transmitted over a single fibre. The two filters used in the experiments feature insertion losses of 0.9dB (100GHz filter) and 2dB (25GHz filter).

manuscript with contributions from the other authors. All authors discussed experiments, results and the interpretation of results.

**Acknowledgements** We thank Andrew Lord and Tim Whitley of BT for useful discussions.

**Author information** The authors declare that they have no competing financial interests. Correspondence and requests for materials should be addressed to J.F.D. (email: james.dynes@crl.toshiba.co.uk).